\documentclass[11pt,oneside,english,reqno]{amsart}
\usepackage[T1]{fontenc}
\usepackage{verbatim}
\usepackage{amsthm}
\usepackage{amstext}
\usepackage{amssymb}
\usepackage{esint}
\usepackage{xargs}[2008/03/08]

\makeatletter
\numberwithin{equation}{section}

\usepackage{undertilde}
\newcommand{\ie}{\textit{i.e.}}

\usepackage{mathpazo}

\makeatother

\usepackage{babel}
\begin{document}
\begin{comment}
Basic Math
\end{comment}
\global\long\def\ga{\alpha}
\global\long\def\gb{\beta}
\global\long\def\ggm{\gamma}
\global\long\def\go{\omega}
\global\long\def\ge{\epsilon}
\global\long\def\gs{\sigma}
\global\long\def\gd{\delta}
\global\long\def\gD{\Delta}
\global\long\def\vph{\varphi}
\global\long\def\gf{\varphi}
\global\long\def\gk{\kappa}
\global\long\def\eps{\varepsilon}
\global\long\def\epss#1#2{\varepsilon_{#2}^{#1}}
\global\long\def\ep#1{\eps_{#1}}
\global\long\def\wh#1{\widehat{#1}}
\global\long\def\spec#1{\textsf{#1}}
\global\long\def\ui{\wh{\boldsymbol{\imath}}}
\global\long\def\uj{\wh{\boldsymbol{\jmath}}}
\global\long\def\uk{\widehat{\boldsymbol{k}}}
\global\long\def\uI{\widehat{\mathbf{I}}}
\global\long\def\uJ{\widehat{\mathbf{J}}}
\global\long\def\uK{\widehat{\mathbf{K}}}
\global\long\def\bs#1{\boldsymbol{#1}}
\global\long\def\vect#1{\mathbf{#1}}
\global\long\def\bi#1{\textbf{\emph{#1}}}
\global\long\def\uv#1{\widehat{\boldsymbol{#1}}}
\global\long\def\cross{\times}
\global\long\def\ddt{\frac{\dee}{\dee t}}
\global\long\def\dbyd#1{\frac{\dee}{\dee#1}}
\global\long\def\dby#1#2{\frac{\partial#1}{\partial#2}}
\global\long\def\vct#1{\mathbf{#1}}
\begin{comment}
General Math
\end{comment}
\global\long\def\partialby#1#2{\frac{\partial#1}{\partial x^{#2}}}
\newcommandx\parder[2][usedefault, addprefix=\global, 1=]{\frac{\partial#2}{\partial#1}}
\global\long\def\oneto{1,\dots,}
\global\long\def\mi#1{\boldsymbol{#1}}
\global\long\def\mii{\mi I}
\begin{comment}
Multi-index
\end{comment}
\global\long\def\fall{,\quad\text{for all}\quad}
\global\long\def\reals{\mathbb{R}}
\global\long\def\rthree{\reals^{3}}
\global\long\def\rsix{\reals^{6}}
\global\long\def\rn{\reals^{n}}
\global\long\def\rt#1{\reals^{#1}}
\global\long\def\les{\leqslant}
\global\long\def\ges{\geqslant}
\global\long\def\dee{\textrm{d}}
\global\long\def\di{d}
\global\long\def\from{\colon}
\global\long\def\tto{\longrightarrow}
\global\long\def\abs#1{\left|#1\right|}
\global\long\def\isom{\cong}
\global\long\def\comp{\circ}
\global\long\def\cl#1{\overline{#1}}
\global\long\def\fun{\varphi}
\global\long\def\interior{\textrm{Int}\,}
\global\long\def\sign{\textrm{sign}\,}
\global\long\def\sgn#1{(-1)^{#1}}
\global\long\def\sgnp#1{(-1){}^{\abs{#1}}}
\global\long\def\dimension{\textrm{dim}\,}
\global\long\def\esssup{\textrm{ess}\,\sup}
\global\long\def\ess{\textrm{{ess}}}
\global\long\def\kernel{\mathop{\textrm{Kernel}}}
\global\long\def\support{\textrm{supp}\,}
\global\long\def\image{\textrm{Image}\,}
\global\long\def\diver{\mathop{\textrm{div}}}
\global\long\def\sp{\mathop{\textrm{span}}}
\global\long\def\resto#1{|_{#1}}
\global\long\def\incl{\iota}
\global\long\def\iden{\imath}
\global\long\def\idnt{\textrm{Id}}
\global\long\def\rest{\rho}
\global\long\def\extnd{e_{0}}
\global\long\def\proj{\textrm{pr}}
\global\long\def\ino#1{\int_{#1}}
\global\long\def\half{\frac{1}{2}}
\global\long\def\shalf{{\scriptstyle \half}}
\global\long\def\third{\frac{1}{3}}
\global\long\def\empt{\varnothing}
\global\long\def\paren#1{\left(#1\right)}
\global\long\def\bigp#1{\bigl(#1\bigr)}
\global\long\def\biggp#1{\biggl(#1\biggr)}
\global\long\def\Bigp#1{\Bigl(#1\Bigr)}
\global\long\def\braces#1{\left\{  #1\right\}  }
\global\long\def\sqbr#1{\left[#1\right]}
\global\long\def\anglep#1{\left\langle #1\right\rangle }
\global\long\def\lsum{{\textstyle \sum}}
\global\long\def\bigabs#1{\bigl|#1\bigr|}
\global\long\def\lisub#1#2#3{#1_{1}#2\dots#2#1_{#3}}
\global\long\def\lisup#1#2#3{#1^{1}#2\dots#2#1^{#3}}
\global\long\def\lisubb#1#2#3#4{#1_{#2}#3\dots#3#1_{#4}}
\global\long\def\lisubbc#1#2#3#4{#1_{#2}#3\cdots#3#1_{#4}}
\global\long\def\lisubbwout#1#2#3#4#5{#1_{#2}#3\dots#3\widehat{#1}_{#5}#3\dots#3#1_{#4}}
\global\long\def\lisubc#1#2#3{#1_{1}#2\cdots#2#1_{#3}}
\global\long\def\lisupc#1#2#3{#1^{1}#2\cdots#2#1^{#3}}
\global\long\def\lisupp#1#2#3#4{#1^{#2}#3\dots#3#1^{#4}}
\global\long\def\lisuppc#1#2#3#4{#1^{#2}#3\cdots#3#1^{#4}}
\global\long\def\lisuppwout#1#2#3#4#5#6{#1^{#2}#3#4#3\wh{#1^{#6}}#3#4#3#1^{#5}}
\global\long\def\lisubbwout#1#2#3#4#5#6{#1_{#2}#3#4#3\wh{#1}_{#6}#3#4#3#1_{#5}}
\global\long\def\lisubwout#1#2#3#4{#1_{1}#2\dots#2\widehat{#1}_{#4}#2\dots#2#1_{#3}}
\global\long\def\lisupwout#1#2#3#4{#1^{1}#2\dots#2\widehat{#1^{#4}}#2\dots#2#1^{#3}}
\global\long\def\lisubwoutc#1#2#3#4{#1_{1}#2\cdots#2\widehat{#1}_{#4}#2\cdots#2#1_{#3}}
\global\long\def\twp#1#2#3{\dee#1^{#2}\wedge\dee#1^{#3}}
\global\long\def\thp#1#2#3#4{\dee#1^{#2}\wedge\dee#1^{#3}\wedge\dee#1^{#4}}
\global\long\def\fop#1#2#3#4#5{\dee#1^{#2}\wedge\dee#1^{#3}\wedge\dee#1^{#4}\wedge\dee#1^{#5}}
\global\long\def\idots#1{#1\dots#1}
\global\long\def\icdots#1{#1\cdots#1}
\global\long\def\norm#1{\|#1\|}
\global\long\def\nonh{\heartsuit}
\global\long\def\nhn#1{\norm{#1}^{\nonh}}
\global\long\def\trps{^{{\scriptscriptstyle \textsf{T}}}}
\global\long\def\testfuns{\mathcal{D}}
\global\long\def\ntil#1{\tilde{#1}{}}
\begin{comment}
Forms-Differential Geometry
\end{comment}
\global\long\def\alt{\mathfrak{A}}
\global\long\def\pou{\eta}
\global\long\def\ext{{\textstyle \bigwedge}}
\global\long\def\forms{\Omega}
\global\long\def\dotwedge{\dot{\mbox{\ensuremath{\wedge}}}}
\global\long\def\vel{\theta}
\begin{comment}
<volume element
\end{comment}
\global\long\def\contr{\raisebox{0.4pt}{\mbox{\ensuremath{\lrcorner}}}\,}
\global\long\def\lie{\mathcal{L}}
\global\long\def\L#1{L\bigl(#1\bigr)}
\global\long\def\vvforms{\ext^{\dims}\bigp{T\spc,\vbts^{*}}}
\begin{comment}
>\textcompwordmark{}>\textcompwordmark{}>\textcompwordmark{}>\textcompwordmark{}>\textcompwordmark{}>Space
Time Events<\textcompwordmark{}<\textcompwordmark{}<\textcompwordmark{}<\textcompwordmark{}<\textcompwordmark{}<\textcompwordmark{}<\textcompwordmark{}<
\end{comment}
\global\long\def\spc{\mathcal{S}}
\global\long\def\sptm{\mathcal{E}}
\global\long\def\evnt{e}
\global\long\def\frame{\Phi}
\global\long\def\timeman{\mathcal{T}}
\global\long\def\zman{t}
\global\long\def\dims{n}
\global\long\def\m{\dims-1}
\global\long\def\dimw{m}
\global\long\def\wc{z}
\global\long\def\fourv#1{\mbox{\ensuremath{\mathfrak{#1}}}}
\global\long\def\pbform#1{\utilde{#1}}
\global\long\def\util#1{\raisebox{-5pt}{\ensuremath{{\scriptscriptstyle \sim}}}\!\!\!#1}
\global\long\def\utilJ{\util J}
\global\long\def\utilRho{\util{\rho}}
\global\long\def\body{B}
\global\long\def\bdry{\partial}
\global\long\def\gO{\varOmega}
\global\long\def\reg{\mathcal{R}}
\global\long\def\bdrr{\bdry\reg}
\global\long\def\bdom{\bdry\gO}
\global\long\def\bndo{\partial\gO}
\begin{comment}
{*}{*}{*}{*}{*}{*}{*}{*}{*}{*}{*}{*}{*}{*}{*}{*}{*}{*}{*}{*}{*}{*}{*}{*}{*}{*}{*}{*}{*}{*}{*}{*}{*}{*}{*}{*}{*}{*}{*}{*}{*}{*}
{*}{*}{*}{*}{*}{*}{*}{*}{*}{*}{*}{*}{*}{*}{*}{*}{*}{*}{*}Cauchy Fluxes{*}{*}{*}{*}{*}{*}{*}{*}{*}{*}
{*}{*}{*}{*}{*}{*}{*}{*}{*}{*}{*}{*}{*}{*}{*}{*}{*}{*}{*}{*}{*}{*}{*}{*}{*}{*}{*}{*}{*}{*}{*}{*}{*}{*}{*}{*}{*}{*}{*}{*}{*}{*}{*}
\end{comment}
\global\long\def\pform{\varsigma}
\global\long\def\vform{\beta}
\global\long\def\sform{\tau}
\global\long\def\flow{J}
\global\long\def\n{\m}
\global\long\def\cmap{\mathfrak{t}}
\global\long\def\vcmap{\varSigma}
\global\long\def\mvec{\mathfrak{v}}
\global\long\def\mveco#1{\mathfrak{#1}}
\global\long\def\smbase{\mathfrak{e}}
\global\long\def\spx{\simp}
\global\long\def\hp{H}
\global\long\def\ohp{h}
\global\long\def\hps{G_{\dims-1}(T\spc)}
\global\long\def\ohps{G_{\dims-1}^{\perp}(T\spc)}
\global\long\def\hpsx{G_{\dims-1}(\tspc)}
\global\long\def\ohpsx{G_{\dims-1}^{\perp}(\tspc)}
\global\long\def\fbun{F}
\global\long\def\flowm{\Phi}
\global\long\def\tgb{T\spc}
\global\long\def\ctgb{T^{*}\spc}
\global\long\def\tspc{T_{\pis}\spc}
\global\long\def\dspc{T_{\pis}^{*}\spc}
\begin{comment}
{*}{*}{*}{*}{*} ELECTROMAGNETISM IN SPACETIME <\textcompwordmark{}<\textcompwordmark{}<\textcompwordmark{}<\textcompwordmark{}<\textcompwordmark{}<
\end{comment}
%%%%%%% ELECTROMAGNETISM IN SPACETIME %%%%%%
\global\long\def\fflow{\fourv J}
%      four-flow
\global\long\def\fvform{\mathfrak{b}}
%      four body flux
\global\long\def\fsform{\mathfrak{t}}
%      four surface flux
\global\long\def\fpform{\mathfrak{s}}
%      Four production rate
%\newcommand{\fgr}{\mathfrak{I}}%         Growth rate in space time
\global\long\def\maxw{\mathfrak{g}}
%        Maxwell form or Stream form
\global\long\def\frdy{\mathfrak{f}}
%        Faraday 2-form
%\newcommand{\maxw}{\mathcal{G}}%        Maxwell form or Stream form
%\newcommand{\frdy}{\mathcal{F}}%        Faraday 2-form
\global\long\def\ptnl{A}
%                   Potential 1-form
%%%%%%%%%%%%%%%%%%%%%%%%%%%%%%%%%%%%%%%%%%%%%
\begin{comment}
{*}{*}{*}{*}{*} Jets and Vector Bundles {*}{*}{*}{*}{*}
\end{comment}
\global\long\def\eucl{E}
\global\long\def\mind{\alpha}
\global\long\def\vb{\xi}
\global\long\def\man{\mathcal{M}}
\global\long\def\odman{\mathcal{N}}
\global\long\def\subman{\mathcal{A}}
\global\long\def\vbt{\mathcal{E}}
\global\long\def\fib{\mathbf{V}}
\global\long\def\vbts{W}
\global\long\def\avb{U}
\global\long\def\chart{\varphi}
\global\long\def\vbchart{\Phi}
\global\long\def\jetb#1{J^{#1}}
\global\long\def\jet#1{j^{1}(#1)}
\global\long\def\Jet#1{J^{1}(#1)}
\global\long\def\jetm#1{j_{#1}}
\begin{comment}
Sobolev Spaces
\end{comment}
\global\long\def\sobp#1#2{W_{#2}^{#1}}
\global\long\def\inner#1#2{\left\langle #1,#2\right\rangle }
\global\long\def\fields{\sobp pk(\vb)}
\global\long\def\bodyfields{\sobp p{k_{\partial}}(\vb)}
\global\long\def\forces{\sobp pk(\vb)^{*}}
\global\long\def\bfields{\sobp p{k_{\partial}}(\vb\resto{\bndo})}
\global\long\def\loadp{(\sfc,\bfc)}
\global\long\def\strains{\lp p(\jetb k(\vb))}
\global\long\def\stresses{\lp{p'}(\jetb k(\vb)^{*})}
\global\long\def\diffop{D}
\global\long\def\strainm{E}
\global\long\def\incomps{\vbts_{\yieldf}}
\global\long\def\devs{L^{p'}(\eta_{1}^{*})}
\global\long\def\incompsns{L^{p}(\eta_{1})}
\begin{comment}
Distributions and Currents
\end{comment}
\global\long\def\testf{\mathcal{D}}
\global\long\def\dists{\mathcal{D}'}
\global\long\def\codiv{\boldsymbol{\partial}}
\global\long\def\currof#1{\tilde{#1}}
\global\long\def\chn{c}
\global\long\def\chnsp{\mathbf{F}}
\global\long\def\current{T}
\global\long\def\curr#1{T_{\langle#1\rangle}}
\global\long\def\prop{P}
\global\long\def\aprop{Q}
\global\long\def\flux{T}
\global\long\def\aflux{S}
\global\long\def\fform{\tau}
\global\long\def\dimn{n}
\global\long\def\sdim{{\dimn-1}}
\global\long\def\contrf{{\scriptstyle \smallfrown}}
\global\long\def\prodf{{\scriptstyle \smallsmile}}
\global\long\def\ptnl{\varphi}
\global\long\def\form{\omega}
\global\long\def\dens{\rho}
\global\long\def\simp{s}
\global\long\def\ssimp{\Delta}
\global\long\def\cpx{K}
\global\long\def\cell{C}
\global\long\def\chain{B}
\global\long\def\ach{A}
\global\long\def\coch{X}
\global\long\def\scale{s}
\global\long\def\fnorm#1{\norm{#1}^{\flat}}
\global\long\def\chains{\mathcal{A}}
\global\long\def\ivs{\boldsymbol{U}}
\global\long\def\mvs{\boldsymbol{V}}
\global\long\def\cvs{\boldsymbol{W}}
\begin{comment}
Points Vectors and Regions
\end{comment}
\global\long\def\pis{x}
\global\long\def\xo{\pis_{0}}
\global\long\def\pib{X}
\global\long\def\pbndo{\Gamma}
\global\long\def\bndoo{\pbndo_{0}}
 \global\long\def\bndot{\pbndo_{t}}
\global\long\def\cloo{\cl{\gO}}
\global\long\def\nor{\mathbf{n}}
\global\long\def\dA{\,\dee A}
\global\long\def\dV{\,\dee V}
\global\long\def\eps{\varepsilon}
\global\long\def\vs{\mathbf{W}}
\global\long\def\avs{\mathbf{V}}
\global\long\def\affsp{\mathbf{A}}
\global\long\def\pt{p}
\global\long\def\vbase{e}
\global\long\def\sbase{\mathbf{e}}
\global\long\def\msbase{\mathfrak{e}}
\global\long\def\vect{v}
\begin{comment}
Kinematics, Strains
\end{comment}
\global\long\def\vf{w}
\global\long\def\avf{u}
\global\long\def\stn{\varepsilon}
\global\long\def\rig{r}
\global\long\def\rigs{\mathcal{R}}
\global\long\def\qrigs{\!/\!\rigs}
\global\long\def\qd{\!/\,\!\kernel\diffop}
\global\long\def\dis{\chi}
\global\long\def\conf{\kappa}
\begin{comment}
Forces and Stresses
\end{comment}
\global\long\def\fc{F}
\global\long\def\st{\sigma}
\global\long\def\bfc{\mathbf{b}}
\global\long\def\sfc{\mathbf{t}}
\global\long\def\stm{S}
\global\long\def\nhs{Y}
\begin{comment}
Nonholonomic Variational Second Order Stress
\end{comment}
\global\long\def\soc{Z}
\begin{comment}
Second Order Cauchy Stress
\end{comment}
\global\long\def\tran{\mathrm{tr}}
\global\long\def\slf{R}
%Self force
\global\long\def\sts{\varSigma}
%stresses
\global\long\def\ebdfc{T}
\global\long\def\optimum{\st^{\textrm{opt}}}
\global\long\def\scf{K}
\begin{comment}
Function Spaces
\end{comment}
\global\long\def\cee#1{C^{#1}}
\global\long\def\lone{L^{1}}
\global\long\def\linf{L^{\infty}}
\global\long\def\lp#1{L^{#1}}
\global\long\def\ofbdo{(\bndo)}
\global\long\def\ofclo{(\cloo)}
\global\long\def\vono{(\gO,\rthree)}
\global\long\def\vonbdo{(\bndo,\rthree)}
\global\long\def\vonbdoo{(\bndoo,\rthree)}
\global\long\def\vonbdot{(\bndot,\rthree)}
\global\long\def\vonclo{(\cl{\gO},\rthree)}
\global\long\def\strono{(\gO,\reals^{6})}
\global\long\def\sob{W_{1}^{1}}
\global\long\def\sobb{\sob(\gO,\rthree)}
\global\long\def\lob{\lone(\gO,\rthree)}
\global\long\def\lib{\linf(\gO,\reals^{12})}
\global\long\def\ofO{(\gO)}
\global\long\def\oneo{{1,\gO}}
\global\long\def\onebdo{{1,\bndo}}
\global\long\def\info{{\infty,\gO}}
\global\long\def\infclo{{\infty,\cloo}}
\global\long\def\infbdo{{\infty,\bndo}}
\global\long\def\ld{LD}
\global\long\def\ldo{\ld\ofO}
\global\long\def\ldoo{\ldo_{0}}
\global\long\def\trace{\gamma}
\global\long\def\pr{\proj_{\rigs}}
\global\long\def\pq{\proj}
\global\long\def\qr{\,/\,\reals}
\begin{comment}
Plasticity and Optimization
\end{comment}
\global\long\def\aro{S_{1}}
\global\long\def\art{S_{2}}
\global\long\def\mo{m_{1}}
\global\long\def\mt{m_{2}}
\begin{comment}
Optimization
\end{comment}
\global\long\def\yieldc{B}
\global\long\def\yieldf{Y}
\global\long\def\trpr{\pi_{P}}
\global\long\def\devpr{\pi_{\devsp}}
\global\long\def\prsp{P}
\global\long\def\devsp{D}
\global\long\def\ynorm#1{\|#1\|_{\yieldf}}
\global\long\def\colls{\Psi}
%Collapse sufrace
\begin{comment}
Finite Elements
\end{comment}
\global\long\def\ssx{S}
\global\long\def\smap{s}
\global\long\def\smat{\chi}
\global\long\def\sx{e}
\global\long\def\snode{P}
\global\long\def\elem{e}
\global\long\def\nel{L}
\global\long\def\el{l}
\global\long\def\ipln{\phi}
\global\long\def\ndof{D}
\global\long\def\dof{d}
\global\long\def\nldof{N}
\global\long\def\ldof{n}
\global\long\def\lvf{\chi}
\global\long\def\lfc{\varphi}
\global\long\def\amat{A}
\global\long\def\snomat{E}
\global\long\def\femat{E}
\global\long\def\tmat{T}
\global\long\def\fvec{f}
\global\long\def\snsp{\mathcal{S}}
\global\long\def\slnsp{\Phi}
\global\long\def\ro{r_{1}}
\global\long\def\rtwo{r_{2}}
\global\long\def\rth{r_{3}}
\global\long\def\subbs{\mathcal{B}}
\global\long\def\elements{\mathcal{E}}
\global\long\def\element{E}
\global\long\def\nodes{\mathcal{N}}
\global\long\def\node{N}
\global\long\def\psubbs{\mathcal{P}}
\global\long\def\psubb{P}
\global\long\def\matr{M}
\global\long\def\nodemap{\nu}
\begin{comment}
{*}{*}{*}{*}{*}{*}
FINITE CHAINS
{*}{*}{*}{*}{*}{*}{*}{*}
\end{comment}
\global\long\def\node{v}
\global\long\def\edge{e}
\global\long\def\accu{q}
\global\long\def\accusp{\mathcal{Q}}
\global\long\def\potl{\varphi}
\global\long\def\ptnl{\alpha}
\global\long\def\currsp{\mathcal{I}}
\global\long\def\volt{V}
\global\long\def\intv{\mathbf{t}}
\global\long\def\intc{t}
\global\long\def\intsp{\mathcal{T}}
\global\long\def\frcv{\mathbf{f}}
\global\long\def\frcc{f}
\global\long\def\frcsp{\mathcal{F}}
\global\long\def\velv{\mathbf{V}}
\global\long\def\velc{V}
\global\long\def\disv{\mathbf{E}}
\global\long\def\disc{E}
\global\long\def\posn{\mathbf{x}}
\global\long\def\area{\mathbf{A}}
\global\long\def\relp{\mathbf{L}}
\global\long\def\chn{c}

\title[Second Order Stresses]{Metric Independent Analysis of \\Second Order Stresses}

\author{Reuven Segev}

\curraddr{Reuven Segev\\
Department of Mechanical Engineering\\
Ben-Gurion University of the Negev\\
Beer-Sheva, Israel\\
rsegev@bgu.ac.il}

\keywords{Continuum mechanics; high order stresses stress; virtual power; differentiable
manifolds; jet bundles; iterated jets, non-holonomic sections.}

\thanks{\today}

\subjclass[2000]{74A10; 53Z05 }
\begin{abstract}
A metric independent geometric analysis of second order stresses in
continuum mechanics is presented. For a vector bundle $\vbts$ over
the $n$-dimensional space manifold, the value of a second order stress
at a point $x$ in space is represented mathematically by a linear
mapping between the second jet space of $\vbts$ at $x$ and the space
of $n$-alternating tensors at $x$. While only limited analysis can
be performed on second order stresses as such, they may be represented
by non-holonomic stresses, whose values are linear mapping defined
on the iterated jet bundle, $J^{1}(J^{1}\vbts)$, and for which an
iterated analysis for first order stresses may be performed. As expected,
we obtain the surface interactions on the boundaries of regions in
space.
\end{abstract}
\maketitle

\section{Introduction}

A metric independent geometric analysis of second order stresses in
continuum mechanics is presented. A vector bundle $\vbts$ over the
$n$-dimensional space manifold $\spc$ is considered whose sections
are interpreted as virtual generalized velocity fields. In the standard
case, $\vbts=T\spc$, the tangent bundle of $\spc$. A second order
stress $\stm$ is defined to be a tensor field over space such that
the value $\stm(x)$ is a linear mapping between the second jet $J^{2}\vbts_{x}$
space of $\vbts$ at $x$ and the space $\ext^{n}T_{x}^{*}\spc$ of
$n$-alternating tensors at $x$. Thus, for a smooth $n$-dimensional
submanifold $\body$ in space, representing the image of a body, and
a virtual velocity field $\vf$,
\begin{equation}
\int_{\body}S(j^{2}\vf),
\end{equation}
where $j^{2}\vf$ denotes the second jet of $\vf$, represents the
virtual power performed by the stress field.

Only limited analysis can be performed on second order stresses. However,
second order stresses may be represented, non-uniquely, by non-holonomic
stress. By a non-holonomic stress we refer to a tensor field whose
value at a point $x\in\spc$ is a linear mapping from the iterated
jet bundle, $J^{1}(J^{1}\vbts)_{x}$, to $\ext^{2}T_{x}^{*}\spc$.
Non-holonomic stresses may be treated by repeating the metric independent
analysis for first order stresses as in \cite{Segev2002,Segev2013}.
As expected, we obtain the surface interactions on the boundaries
of regions in space.

\section{Notation and Preliminaries\label{sec:Notation-and-Prliminaries}}

We will use the same scheme of notation as in \cite{Segev2013} and
we will ofter use the same notation for a mapping and variables in
the co-domains thereof. Thus, Let $\vb:\vbts\to\spc$ be a vector
bundle. We recall that a section $\vf:\spc\to\vbts$ of $\vb$ is
represented locally in the form
\begin{equation}
(\lisup x,{\dims})\longmapsto(\lisup x,n,\vf^{1}(x^{i}),\dots,\vf^{d}(x^{i}))\label{eq:ReprSectVB-1}
\end{equation}
where $(\lisup x,n)$ is a local coordinate system and a local basis
$\{g_{1},\dots,g_{d}\}$ was used for the fibers of $\vbts$. Let
$\mii=(i_{1},\dots,i_{n})$, for non-negative integers $i_{j}$, be
a multi-index and let $\abs{\mii}=\sum_{j=1}^{n}i_{j}$. We use the
notation
\begin{equation}
\parder[x^{\mii}]{^{\abs{\mii}}}=\parder[x^{i_{1}}\cdots\bdry x^{i_{n}}]{^{\abs{\mii}}}.
\end{equation}
Two sections $\vf$ and $\vf'$ have the same \emph{$k$-jet at $x_{0}\in\spc$}
if
\begin{equation}
\parder[x^{\mii}]{^{\abs{\mii}}\vf^{\ga}}(x_{0}^{i})=\parder[x^{\mii}]{^{\abs{\mii}}\vf'^{\ga}}(x_{0}^{i})
\end{equation}
for all $\mii$ such that $\abs{\mii}\les k$ and all $\ga=1,\dots,d$.
Clearly, if this condition holds in one vector bundle chart in a neighborhood
of $x_{0}$, it will hold using any other chart and it induces an
equivalence relation on the vector space $C^{k}(\xi)$ of $C^{k}$-sections
of the vector bundle. An equivalence class for this relation is \emph{a
$k$-jet at} $x_{0}$. Given a section $\vf$, the jet it induces
at $x_{0}$---the \emph{jet of $w$ at} $x_{0}$---will be denoted
as $j^{k}(w)(x_{0})$. Given a chart in a neighborhood of $x_{0}$,
$j^{k}(\vf)(x_{0})$ is represented by
\begin{equation}
\left\{ \vf_{,\mii}^{\ga}(x_{0}):=\parder[x^{\mii}]{^{\abs{\mii}}\vf^{\ga}}(x_{0}^{i})\mid\abs{\mii}\les k,\,\ga=1,\dots,d\right\} .\label{eq:ReprOfaJet}
\end{equation}

The collection of all $k$-jets of sections at $x_{0}\in\spc$ is
the \emph{$k$-jet space }of the vector bundle at $x$ and is denoted
as $J_{x}^{k}\vbts$. Given a point $x_{0}\in\spc$ and an element
$\vf_{0}\in\vbts_{x_{0}}$, the collection of all $k$-jets at $x_{0}$
such that each jet is represented by a section $w$ with $w(x_{0})=w_{0}$
will be referred to as the \emph{$k$-jet space} at $\vf_{0}$ and
will be denoted by $J_{\vf_{0}}^{k}\vbts$. Evidently,
\begin{equation}
J_{x_{0}}^{k}W=\bigcup_{\vf_{0}\in\vbts_{x_{0}}}J_{\vf_{0}}^{k}\vbts.
\end{equation}
The $k$-\emph{jet bundle} $J^{k}\vbts$ is the collection of all
$k$-jets at the various points in $\spc$ so that
\begin{equation}
J^{k}\vbts=\bigcup_{x\in\spc}J_{x}^{k}\vbts=\bigcup_{\vf\in\vbts}J_{\vf}^{k}\vbts.
\end{equation}
A natural vector bundle structure
\begin{equation}
\pi^{k}:J^{k}\vbts\tto\spc,
\end{equation}
is available on the jet bundle by which $\pi^{k}(A)=x$ if $A\in J_{x}^{k}\vbts$.
The linear structure on the fibers is given by $a_{1}A_{1}+a_{2}A_{2}=j^{k}(a_{1}\vf_{1}+a_{2}\vf_{2})(x)$,
for $A_{1}$, $A_{2}$ in $J_{x}^{k}\vbts$, $a_{1},a_{2}\in\reals$,
and representing sections $\vf_{1}$ and $\vf_{2}$. Evidently, the
result is independent of the choice of representative sections. The
fiber $J_{x}^{k}\vbts$ of this vector bundle over $x\in\spc$ is
isomorphic with
\begin{equation}
\vbts_{x}\oplus\L{T_{x}\spc,\vbts_{x}}\oplus\cdots\oplus L_{S}^{p}(T_{x}\spc,\vbts_{x})\oplus\cdots\oplus L_{S}^{k}(T_{x}\spc,\vbts_{x}),\label{eq:FiberOfJetBundle}
\end{equation}
where $L_{S}^{p}(T_{x}\spc,\vbts_{x})$ denotes the vector space of
$p$-multilinear symmetric mappings from $T_{x}\spc$ to $\vbts_{x}$.
Thus, an element in $J_{x}^{k}\vbts$ is represented locally in the
form
\begin{equation}
(A^{0\ga_{0}},A_{\mii_{1}}^{1\ga_{1}},\dots,A_{\mii_{p}}^{p\ga_{p}},\dots,A_{\mii_{k}}^{k\ga_{k}})=(A_{\mii}^{p\ga}),\label{eq:repJet}
\end{equation}
where $p=0,\dots,k$, $\abs{\mii_{p}}=p$, $\ga=(\ga_{0},\dots,\ga_{k})$,
$\ga_{p}=1,\dots,d$, and evidently, $A^{0\ga_{0}}$ represents an
element of $\vbts_{x}$. Evidently, each section $\vf$ of $\vbts$
induces the section $j^{k}\vf$ of the $k$-th jet bundle and if fact
we have a linear mapping
\begin{equation}
j^{k}:C^{k}(\vbts)\tto C^{0}(J^{k}\vbts),
\end{equation}
where $C^{p}(\avb)$ represents the vector space of sections of the
vector bundle $\avb$ of class $p$. For additional information on
jet bundles, some of which will be used in the following sections
see \cite{Saunders}.

The jet bundle has also a natural projection
\begin{equation}
\pi_{p}^{k}:J^{k}W\tto J^{p}\vbts,\quad0\les p\les k,
\end{equation}
characterized by $\pi_{p}^{k}(A)=j^{p}(\vf)(x)$ where $x=\pi^{k}(A)$
and $\vf$ is any section of $\vbts$ that represents $A$. The mapping
$\pi_{p}^{k}$ is a vector bundle morphism over the identity of $\spc$.

\section{Simple Stresses\label{sec:Simple-Stresses}}

As a primitive mathematical object pertaining to stress theory for
continuum mechanics of order 1 we take the variational stress, a smooth
section $\stm$ of the vector bundle $L(J^{1}\vbts,\ext^{n}T^{*}\spc)$
for some vector bundle $\vbts\to\spc$, where $\ext^{n}T^{*}\spc$
is the vector bundle of $n$-alternating covariant tensors over $\spc$.
For motivation, see \cite{Segev1986,Segev2002,Segev2013}. In particular,
for an $n$-dimensional submanifold with boundary $\body\subset\spc$,
one is interested in the linear functional
\begin{equation}
\vf\longmapsto\int_{\body}\stm(j^{1}\vf)
\end{equation}
which is interpreted as the virtual power performed by the variational
stress $\stm$ for the virtual generalized velocity field $\vf$ inside
the region $\body$.

Locally, $\stm$ is represented locally in the form $(x^{i},R_{1\dots dp},S_{1\dots dp}^{k})$
or equivalently
\begin{equation}
\left(\sum_{\ga}R_{1\dots n\ga}(\lisupc{\dee x}{\wedge}n)\otimes g^{\ga},\sum_{i,\ga}\stm_{1\dots n\ga}^{i}(\lisupc{\dee x}{\wedge}n)\otimes\partialby{}i\otimes g^{\ga}\right)
\end{equation}
so that $\stm(j(\vf))$ is represented locally by
\begin{equation}
\left(\sum_{\ga}R_{1\dots n\ga}\vf^{\ga}+\sum_{i,\ga}\stm_{1\dots n\ga}^{i}\vf_{,i}^{\ga}\right)\lisupc{\dee x}{\wedge}n.\label{eq:ReprPowerIntern-1}
\end{equation}

For a vector bundle $V\to\spc$, let $\ext^{p}(T^{*}\spc,V)$ denote
the vector bundle over $\spc$ whose fiber at $x$ is the vector space
of $p$-alternating multilinear mappings from $T_{x}\spc$ to $V_{x}$.
Consider the isomorphism
\begin{equation}
\tran:\ext^{p}(T^{*}\spc,V^{*})\tto L(V,\ext^{p}T^{*}\spc)
\end{equation}
defined as follows. For $T\in\ext^{p}(T^{*}\spc,V^{*})$, $T^{\tran}=\tran(T)$
is given by
\begin{equation}
T^{\tran}(v)(\lisub u,p)=T(\lisub u,p)(v).
\end{equation}
Thus, for a variational stress $\stm$ one may consider $\stm\trps=\tran^{-1}(\stm)$,
an $n$-form on $\spc$ valued in the dual of the jet bundle.

Using the vector bundle morphism $\pi_{0}^{1}:J\vbts\to\vbts$, consider
the vertical sub-bundle $VJ^{1}\vbts=\kernel\pi_{0}^{1}$. Since $\pi_{0}^{1}$
is represented locally by $(x^{i},u^{p},A_{k}^{q})\mapsto(x^{i},u^{p})$,
an element of $VJ^{1}\vbts$ is represented locally in the form $(x^{i},0,A_{k}^{q})$.
In other words, the fiber $VJ_{x_{0}}^{1}\vbts$ contains jets of
sections of $\vbts$ that vanish at $x_{0}$. It is noted that there
is a natural vector bundle isomorphism $VJ^{1}\vbts\isom\L{T\spc,\vbts}$
by which an element of $VJ^{1}\vbts$ is represented in the form $(x^{i},A_{k}^{q})$.
We use $\incl_{V}:VJ^{1}\vbts\hookrightarrow J^{1}\vbts$ to denote
the inclusion of the vertical sub-bundle---a vector bundle morphism
over $\spc$ represented by $(x^{i},A_{k}^{q})\mapsto(x^{i},0,A_{k}^{q})$.
Then, the dual vector bundle morphism $\incl_{V}^{*}:(J^{1}\vbts)^{*}\to(VJ^{1}\vbts)^{*}\isom\L{\vbts,T\spc}$
is a projection represented locally in the form $(x^{i},\xi_{p},\Xi_{q}^{i})\mapsto(x^{i},\Xi_{q}^{i})$---the
restriction of $\xi\in(J^{1}\vbts)^{*}$ to vertical elements of the
jet bundle. Thus, $\incl_{V}^{*}(\xi)(A)$, $A\in VJ^{1}\vbts$, is
represented by $\sum_{i,q}\Xi_{q}^{i}A_{i}^{q}$. Similarly, for a
section $\stm$ of $\L{J^{1}\vbts,\ext^{n}T^{*}\spc}$, $\incl_{V}^{*}(\stm):=\incl_{V}^{*}\comp\stm$,
a section of $\L{VJ^{1}\vbts,\ext^{d}T^{*}\spc}$, is given by $\incl_{V}^{*}(\stm)(x)(A)=\stm(x)(\incl_{V}(A))\in\ext^{n}T_{x}^{*}\spc$.
The evaluation $\incl_{V}^{*}(\stm)(x)(A)$ is represented by $\sum_{k,\ga}\stm_{1\dots n\ga}^{k}(x)A_{,k}^{\ga}\lisupc{\dee x}{\wedge}n$
and so $\incl_{V}^{*}(\stm)$ is represented in the form w
\begin{equation}
\sum_{k,\ga}\stm_{1\dots n\ga}^{k}(\lisupc{\dee x}{\wedge}n)\otimes\partialby{}k\otimes g^{\ga}.\label{eq:RepresSymbol-1}
\end{equation}
The object $\incl_{V}^{*}(\stm)$ is the symbol of the linear differential
operator $\stm$ as defined in \cite{Palais68}.

Using the isomorphism $VJ^{1}\vbts\isom\L{T\spc,\vbts}$, we regard
$\incl_{V}^{*}(\stm)$ as a section of
\begin{equation}
\begin{split}\L{\L{T\spc,\vbts},\ext^{n}T^{*}\spc} & \isom\bigl(\ext^{n}T^{*}\spc\bigr)\otimes\L{T\spc,\vbts}^{*},\\
 & \isom\bigl(\ext^{n}T^{*}\spc\bigr)\otimes\L{\vbts,T\spc},\\
 & \isom\bigl(\ext^{n}T^{*}\spc\bigr)\otimes T\spc\otimes\vbts^{*}.
\end{split}
\end{equation}
It follows that a section of $\ext^{\dims}(T^{*}\spc,L(\vbts,T\spc))$
may be represented locally in the form $\sum_{a}\theta\otimes v_{a}\otimes\fun^{a}$
for an $n$-form $\theta$ and pairs $v_{a},$ $\fun^{a}$ of sections
of $T\spc$ and $\vbts^{*}$, respectively. We can use the contraction
of the second and first factors in the product to obtain $\sum_{a}(v_{a}\contr\theta)\otimes\fun^{a}$.
Thus, we have a natural mapping
\begin{equation}
\begin{split}\spec C:\L{\L{T\spc,\vbts},\ext^{n}T^{*}\spc} & \tto\ext^{n-1}(T^{*}\spc)\otimes\vbts^{*}\\
 & \,\,\,\,\,\isom\L{\vbts,\ext^{n-1}T^{*}\spc}.
\end{split}
\end{equation}
The mapping $\spec C$ is represented locally by
\begin{multline}
\sum_{k,\ga}\stm_{1\dots n\ga}^{k}(\lisupc{\dee x}{\wedge}n)\otimes\partialby{}k\otimes g^{\ga}\longmapsto\sum_{k,\ga}\stm_{1\dots n\ga}^{k}\partialby{}k\contr(\lisupc{\dee x}{\wedge}n)\otimes g^{\ga},\\
=\sum_{k,\ga}(-1)^{k-1}\stm_{1\dots n\ga}^{k}(\lisupwout{\dee x}{\wedge}nk)\otimes g^{\ga},
\end{multline}
where a superimposed ``hat'' indicates the omission of the associated
term.

The mapping
\begin{equation}
p_{\st}:=\spec C\comp\incl_{V}^{*}:\L{\L{T\spc,\vbts},\ext^{n}T^{*}\spc}\tto\L{\vbts,\ext^{n-1}T^{*}\spc}\label{eq:ProjectionToTractionSt}
\end{equation}
associates a section $\st=p_{\st}\comp\stm$ of $\L{\vbts,\ext^{n-1}T^{*}\spc}$
with any variational stress $\stm$. We refer to a section of $\L{\vbts,\ext^{n-1}T^{*}\spc}$
as a \emph{traction stress}. Such a section is represented locally
in the form $(x^{i},\st_{1\dots\wh k\dots n\ga}(x^{j}))$ or
\begin{equation}
\sum_{k,r}\st_{1\dots\wh k\dots n\ga}(\lisuppwout{\dee x}1{\wedge}{\cdots}nk)\otimes g^{\ga}.\label{eq:LocalRepOfTractStresses-1}
\end{equation}
The transposed, $\st\trps$, is represented by
\begin{equation}
\sum_{k,\ga}\st_{1\dots\wh k\dots n\ga}g^{\ga}\otimes(\lisuppwout{\dee x}1{\wedge}{\cdots}nk)
\end{equation}
and $\st(\vf)$ is represented locally by
\begin{equation}
\sum_{k,\ga}\st_{1\dots\wh k\dots n\ga}\vf^{\ga}\lisuppwout{\dee x}1{\wedge}{\cdots}nk.\label{eq:ReprActionOfTrackStOnVelField-1}
\end{equation}
We conclude that in case $\st=p_{\st}(\stm)$, then,
\begin{equation}
\sigma_{1\dots\wh k\dots n\ga}=(-1)^{k-1}\stm_{1\dots n\ga}^{k}\label{eq:CauchyAndVariational}
\end{equation}
.

For each $(n-1)$-dimensional oriented submanifold $P\subset\spc$,
in particular, the boundary $\bdry\body$ of submanifold with boundary
$\body\subset\spc$, one may integrate $\st(\vf)$ over $P$, and
evaluate
\begin{equation}
\int_{P}\incl_{P}^{*}(\st(\vf)).
\end{equation}
Here, $\incl_{P}:P\to\spc$ is the natural inclusion so that $\incl_{P}^{*}$
is the restriction of forms. We conclude that $\sfc_{P}=\incl_{P}^{*}\comp\st$
is the surface force induced by $\st$ and the integral above represents
the power produced by the traction.

The \emph{divergence}, $\diver\stm$, of the variational stress field
$\stm$ is a section of $\L{\vbts,\ext^{n}(T^{*}\spc)}$ which is
defined invariantly by
\begin{equation}
\diver\stm(\vf)=\dee\paren{p_{\st}(S)(\vf)}-\stm(\jet{\vf)},\label{eq:DefineDivergence-1}
\end{equation}
for every differentiable vector field $\vf$. To present the local
expression for $\diver\stm$ we first note that if $\st=p_{\st}(\stm)$,
then $\dee(\st(\vf))$ is represented locally by

\begin{multline}
\sum_{k,\ga}\dee(\st_{1\dots\wh k\dots n\ga}\vf^{\ga})\wedge\lisupwout{\dee x}{\wedge}nk\\
\begin{split} & =\sum_{k,\ga}(\st_{1\dots\wh k\dots n\ga}\vf^{\ga})_{,k}\dee x^{k}\wedge\lisupwout{\dee x}{\wedge}{\ga}k,\\
 & =\sum_{k,\ga}(\st_{1\dots\wh k\dots n\ga}\vf^{\ga})_{,k}(-1)^{k-1}\lisup{\dee x}{\wedge}n,\\
 & =\sum_{k,\ga}(S_{1\dots n\ga}^{k}\vf^{\ga})_{,k}\lisup{\dee x}{\wedge}n.
\end{split}
\vphantom{}
\end{multline}
Using Equation (\ref{eq:ReprPowerIntern}), the local expression for
$\diver\stm(\vf)$ is therefore
\begin{multline}
\sum_{k,p}\left[(S_{1\dots n\ga}^{k}\vf^{\ga})_{,k}-\left(\sum_{\ga}R_{1\dots n\ga}\vf^{\ga}+\sum_{k,\ga}\stm_{1\dots n\ga}^{k}\vf_{,k}^{\ga}\right)\right]\lisup{\dee x}{\wedge}n\\
=\sum_{k,\ga}(\stm_{1\dots n\ga,k}^{k}-R_{1\dots n\ga})\vf^{\ga}\lisup{\dee x}{\wedge}n
\end{multline}
so that $\diver\stm$ is represented locally by
\begin{equation}
\sum_{k,p}(\stm_{1\dots n\ga,k}^{k}-\slf_{1\dots n\ga})\lisup{\dee x}{\wedge}n\otimes g^{\ga}.\label{eq:ReprDivergence-1}
\end{equation}

It is noted that in the case where $\slf_{1\dots n\ga}=0$ locally,
the expression for the divergence reduces to the traditional expression
for the divergence of a tensor field in a Euclidean space.

Given a variational stress $\stm$, and setting
\begin{equation}
\bfc=-\diver\stm,\label{eq:DiffBalanceLawStresses-1}
\end{equation}
 for every $n$-dimensional submanifold with boundary $\body\subset\spc$,
one has
\begin{equation}
\int_{\body}\bfc(\vf)+\int_{\bdry\body}\sfc_{\bdry\body}(\vf)=\int_{\body}\stm(j(\vf))\label{eq:PrinVirtWorkSimple}
\end{equation}
which is our generalization of the principle of virtual work.

\section{High Order Stresses}

For continuum mechanics of order greater than one, the fundamental
object we consider is the $k$-th order variational stress which is
a smooth section of the vector bundle $L(J^{k}\avb,\ext^{n}T^{*}\spc)\isom\ext^{n}T^{*}\spc\otimes(J^{k}\avb)^{*}$,
for some vector bundle $\avb\to\spc$ whose sections are interpreted
as virtual generalized velocities (see \cite{Segev1986} and \cite{SegevDeBotton1991}
for motivation). Thus, the virtual power performed by the $k$-th
order variational stress $\stm$ for the virtual generalized velocity
$\avf$ in a body $\body\subset\spc$ is given by the action of the
functional
\begin{equation}
\avf\longmapsto\int_{\body}\stm(j^{k}\avf).\label{eq:kthOrdrPower}
\end{equation}
Observing (\ref{eq:FiberOfJetBundle}), it follows that the fiber,
$L(J^{k}\avb,\ext^{n}T^{*}\spc)_{x}$ of $L(J^{k}\avb,\ext^{n}T^{*}\spc)$
at $x\in\spc$ is isomorphic with
\begin{equation}
\ext^{n}T^{*}\spc\otimes\left(\avb_{x}^{*}\oplus\L{T_{x}\spc,\avb_{x}}^{*}\oplus\cdots\oplus L_{S}^{p}(T_{x}\spc,\avb_{x})^{*}\oplus\cdots\oplus L_{S}^{k}(T_{x}\spc,\avb_{x})^{*}\right).
\end{equation}
Let $\stm^{p}$ denote the component of $\stm$ in $\ext^{n}T^{*}\spc\otimes L_{S}^{p}(T_{x}\spc,\vbts_{x})^{*}$.
It follows that the stress may be represented locally in the form
$(\stm^{0},\stm^{1},\dots,\stm^{k})$ and $\stm^{p}$ is represented
locally by an array in the form $\stm_{1\dots n\ga}^{p\mii}$, the
$\mii$ is a multi-index with $\abs{\mii}=p$. The action $\stm(A)$
for an element $A\in J^{k}\avb$ is given by
\begin{equation}
\sum_{\abs{\mii}\les k,\,\ga}\stm_{1\dots n\ga}^{p\mii}A_{\mii}^{p\ga}\dee x^{1}\wedge\cdots\wedge\dee x^{n},
\end{equation}
It is our objective to represent the virtual power for high order
stresses (\ref{eq:kthOrdrPower}) in a form analogous to (\ref{eq:PrinVirtWorkSimple}).
We will concentrate on the case $k=2$.

\section{The Iterated Jet Bundle\label{sec:Iterated-Jet-Bundle}}

Since for any vector bundle $\avb$, $\vbts=J^{1}\avb\to\spc$ is
a vector bundle, also, one may consider the vector bundle $J^{1}\vbts=J^{1}(J^{1}U)\to\spc$.
As any section $A$ of $J^{1}\avb$ is locally of the form $(\avf^{\ga}(x),A_{i}^{\gb}(x))=(A^{0\ga_{0}}(x),A_{i}^{1\ga_{1}}(x))$,
an element of the iterated jet bundle is of the form $(x^{i},B^{0\ga_{0}},B_{i_{1}}^{1\ga_{1}},B_{i_{2}}^{2\ga_{2}},B_{i_{3}i_{4}}^{3\ga_{3}})$.
It is observed that no ``compatibility'', or ``holonomicity'',
is imposed, and for example, $B_{i_{3}i_{4}}^{3\ga_{3}}$ need not
be symmetric in the $i_{3},\, i_{4}$ indices.

It follows that an element $X$ of the dual, $J^{1}(J^{1}U)^{*}$,
of the iterated jet bundle is represented in the form
\begin{equation}
(x^{i},X_{\ga_{0}}^{0},X_{\ga_{1}}^{1i_{1}},X_{\ga_{2}}^{2i_{2}},X_{\ga_{3}}^{3i_{3}i_{4}})
\end{equation}
 whose action on $A\in J^{1}(J^{1}U)_{x}$ is given by
\begin{equation}
\sum_{\ga,i,j}X_{\ga}^{0}B^{0\ga}+X_{\ga}^{1i}B_{i}^{1\ga}+X_{\ga}^{2i}B_{i}^{2\ga}+X_{\ga}^{3ij}B_{ij}^{3\ga}
\end{equation}

It is noted that there is a natural vector bundle inclusion $\incl:J^{2}\avb\to J^{1}(J^{1}\avb)$
such that $\incl_{x}:J^{2}\avb_{x}\to J^{1}(J^{1}\avb)_{x}$ is given
as follows. Let $\avf$ be a section of $\avb$ the represents and
element $B\in J^{2}\avb_{x}$. Then, $\vf=j^{1}\avf$ is a section
of $J^{1}\avb$ whose jet is the target element in $J^{1}(J^{1}\avb)$.
Thus,
\begin{equation}
\incl_{x}:(\avf^{\ga_{0}},\avf_{,i_{1}}^{\ga_{1}},\avf_{,i_{1}i_{2}}^{\ga_{2}})\longmapsto(\avf^{\ga_{0}},\avf_{,i_{1}}^{\ga_{1}},\avf_{,i_{2}}^{\ga_{2}},\avf_{,i_{3}i_{4}}^{\ga_{3}}).
\end{equation}
Evidently, the result is independent of the section chosen and locally
the inclusion is in the form
\begin{equation}
(A^{0\ga_{0}},A_{i_{1}}^{1\ga_{1}},A_{i_{1}i_{2}}^{2\ga_{2}})\longmapsto(A^{0\ga_{0}},A_{i_{1}}^{1\ga_{1}},A_{i_{2}}^{1\ga_{2}},A_{i_{3}i_{4}}^{2\ga_{3}}).
\end{equation}
Thus, the image of $\incl$ contains elements for which the second
and third groups of components are identical and the components in
the fourth group are symmetric.

Since the inclusion $\incl$ is injective, the dual $\incl^{*}:J^{1}((J^{1}U))^{*}\to(J^{2}U)^{*}$
is therefore surjective and it satisfies
\begin{equation}
\begin{split}\incl^{*}(X)(A) & =X(\incl A)\\
 & =\sum_{\ga}X_{\ga}^{0}A^{0\ga}+\sum_{\ga,i}(X_{\ga}^{1i}+X_{\ga}^{2i})A_{i}^{1\ga}+\sum_{\ga}X_{\ga}^{3ij}A_{ij}^{2\ga},\\
 & =\sum_{\ga}X_{\ga}^{0}A^{0\ga}+\sum_{\ga,i}(X_{\ga}^{1i}+X_{\ga}^{2i})A_{i}^{1\ga}+\sum_{\ga}\shalf(X_{\ga}^{3ij}+X_{\ga}^{3ji})A_{ij}^{2\ga}.
\end{split}
\end{equation}
The dual is therefore a restriction represented locally by
\begin{equation}
(x^{i},X_{\ga_{0}}^{0},X_{\ga_{1}}^{1i_{1}},X_{\ga_{2}}^{2i_{2}},X_{\ga_{3}}^{3i_{3}i_{4}})\longmapsto(x^{i},X_{\ga_{0}}^{0},X_{\ga_{1}}^{1i_{1}}+X_{\ga_{1}}^{2i_{1}},\shalf(X_{\ga_{3}}^{3i_{3}i_{4}}+X_{\ga_{3}}^{3i_{4}i_{3}}))
\end{equation}
and being surjective, every second order stress $\stm$ is of the
form $\stm=\incl^{*}X$ for some non-unique section $X$ of $J^{1}(J^{1}U$).
Thus, whatever properties we deduce for elements of $J^{1}(J^{1}\avb)$
will hold for their restriction to $\image\incl$.

It is observed finally (see \cite[p.~169]{Saunders}) that there is
no natural inverse to $\incl$, \ie, a projection $J^{1}(J^{1}\avb)\to J^{2}\avb$.

\section{Non-Holonomic Stresses\label{sec:Non-Holonomic-Stresses}}

As elements of $(J^{1}(J^{1}\avb))^{*}$ represent elements of ($J^{2}U)^{*}$
using the surjective mapping $\incl^{*}$, every second order stress
$\stm$, a section of $L(J^{2}\avb,\ext^{n}T^{*}\spc)\isom\ext^{n}T^{*}\spc\otimes(J^{2}\avb)^{*}$,
may be represented by a section $\nhs$ of $L(J^{1}(J^{1}\avb),\ext^{n}T^{*}\spc)\isom\ext^{n}T^{*}\spc\otimes(J^{1}(J^{1}\avb))^{*}$
in the form $S(j^{2}\avf)=\nhs(\incl\comp j^{2}\avf)$. We will refer
to such a section $\nhs$ as a \emph{non-holonomic stress}. The second
order stress $\stm$ induced by the non-holonomic stress $\nhs$ may
therefore be written as
\begin{equation}
\stm=\incl^{*}\nhs.\label{eq:stressInducedByNHS}
\end{equation}
We thus concentrate our attention in this section to analysis of the
action of non-holonomic stresses in the form
\begin{equation}
\avf\longmapsto\int_{\body}\nhs(j^{1}(A))\label{eq:ActionNHSonJet}
\end{equation}
for a section $A$ of $J^{1}\avb$, and in particular, the compatible
case where $A=j^{1}\avf$, for a section $\avf$ of $\avb$.

Since $J^{1}U\to\spc$ is a vector bundle, we may apply to it all
the analysis described in Section \ref{sec:Simple-Stresses} by substituting
$\vbts=J^{1}\avb$. Using the notation of Section \ref{sec:Iterated-Jet-Bundle},
a non-holonomic stress may be represented locally in the form
\begin{multline}
(\lisupc{\dee x}{\wedge}n)\otimes\left(\lsum_{\ga}Y_{1\dots n\ga}^{0}g^{\ga},\lsum_{\ga,i}Y_{1\dots n\ga}^{1i}\parder[x^{i}]{}\right.\otimes g^{\ga},\\
\lsum_{\ga,i}Y_{1\dots n\ga}^{2i}\parder[x^{i}]{}\otimes g^{\ga},\lsum_{\ga,i,j}Y_{1\dots n\ga}^{3ij}\parder[x^{i}]{}\otimes\left.\parder[x^{j}]{}\otimes g^{\ga}\right),
\end{multline}
where we have omitted the indication of the dependence of the various
fields on $x\in\spc$. Thus, the action $\nhs(B)$. $B\in J^{1}(J^{1}\avb)$
is given by
\begin{equation}
\left(\sum_{\ga,i,j}(\nhs_{1\dots n\ga}^{0}B^{0\ga}+\nhs_{1\dots n\ga}^{1i}B_{i}^{1\ga}+\nhs_{1\dots n\ga}^{2i}B_{i}^{2\ga}+\nhs_{1\dots n\ga}^{3ij}B_{ij}^{3\ga})\right)\lisupc{\dee x}{\wedge}n
\end{equation}
and for the jet $j^{1}A$ of a section $A$ of $J^{1}\avb$,
\begin{equation}
\left(\sum_{\ga,i,j}(\nhs_{1\dots n\ga}^{0}A^{0\ga}+\nhs_{1\dots n\ga}^{1i}A_{i}^{1\ga}+\nhs_{1\dots n\ga}^{2i}A_{,i}^{0\ga}+\nhs_{1\dots n\ga}^{3ji}A_{j,i}^{1\ga})\right)\lisupc{\dee x}{\wedge}n.
\end{equation}
It is observed that in the last expressions the components $Y_{1\dots n\ga}^{0}$
and $\nhs_{1\dots n\ga}^{1i}$ assume the roles of the components
$R_{1\dots n\ga}$ in (\ref{eq:ReprPowerIntern-1}) and the components
$\nhs_{1\dots n\ga}^{2i}$ together with $\nhs_{1\dots n\ga}^{3ji}$
assume the roles of $\stm_{1\dots n\ga}^{i}$.

Using the operator $p_{\st}$ as in (\ref{eq:ProjectionToTractionSt}),
one can extract a section $\soc=p_{\st}\nhs$ of the vector bundle
$L(J^{1}\avb,\ext^{n-1}T^{*}\spc)$. The local representation of $\soc$
is of the form
\begin{equation}
\sum_{k,\ga}(\lisuppwout{\dee x}1{\wedge}{\cdots}nk)\otimes\left(\soc_{1\dots\wh k\dots n\ga}^{0}\otimes g^{\ga},\soc_{1\dots\wh k\dots n\ga}^{1i}\parder[x^{i}]{}\otimes g^{\ga}\right)
\end{equation}
From (\ref{eq:CauchyAndVariational}) it follows that
\begin{equation}
\soc_{1\dots\wh k\dots n\ga}^{0}=(-1)^{k-1}Y_{1\dots n\ga}^{2k},\qquad\soc_{1\dots\wh k\dots n\ga}^{1i}=(-1)^{k-1}Y_{1\dots n\ga}^{3ik}.
\end{equation}

Equation (\ref{eq:DefineDivergence-1}) assumes the form
\begin{equation}
\diver\nhs(A)=\dee\paren{p_{\st}(\nhs)(A)}-\nhs(j^{1}A),
\end{equation}
in which $\diver\nhs$, a section of $\L{J^{1}\avb,\ext^{n}(T^{*}\spc)}$
is represented locally by
\begin{equation}
\lisup{\dee x}{\wedge}n\otimes\sum_{\ga,i,j}\left(Y_{1\dots n\ga,j}^{2j}-\nhs_{1\dots n\ga}^{0}+(Y_{1\dots n\ga,j}^{3ij}-\nhs_{1\dots n\ga}^{1i})\parder[x^{i}]{}\right)\otimes g^{\ga}.
\end{equation}

One conclude that for $\soc=p_{\st}(\nhs)$,
\begin{equation}
\int_{\body}\nhs(j^{1}A)=\int_{\body}\dee(\soc(A))-\int_{\body}\diver\nhs(A)=\int_{\bdry\body}\soc(A)-\int_{\body}\diver\nhs(A).
\end{equation}

For the case where $A=j^{1}\avf$ for a section $\avf$ of $\avb$,
\begin{equation}
\nhs(\jet{j^{1}\avf)}=\dee\paren{p_{\st}(\nhs)(j^{1}\avf)}-\diver\nhs(j^{1}\avf),
\end{equation}
and
\begin{equation}
\int_{\body}\nhs(j^{1}(j^{1}\avf))=\int_{\body}\dee(\soc(j^{1}\avf))-\int_{\body}\diver\nhs(j^{1}\avf)=\int_{\bdry\body}\soc(j^{1}\avf)-\int_{\body}\diver\nhs(j^{1}\avf).
\end{equation}
Similarly to a simple variational stress, $\diver\nhs$ is a section
of $\L{J^{1}\avb,\ext^{n}(T^{*}\spc)}$, we may apply the definition
of the generalized divergence (\ref{eq:DefineDivergence-1}) to it
(substituting $\diver\nhs$ for $\stm$) and so
\begin{equation}
\diver\nhs(j^{1}\avf)=\dee(p_{\st}(\diver\nhs)(\avf))-\diver(\diver\nhs)(\avf).
\end{equation}
Here, similarly to a traction stress $p_{\st}(\diver\nhs)$ is a section
of $\L{\avb,\ext^{n-1}T^{*}\spc}$ and $\diver(\diver\nhs)$ is a
section of $\L{\avb,\ext^{n}(T^{*}\spc)}$, similarly to a body force.
Thus,
\begin{equation}
\begin{split}\int_{\body}\nhs(j^{1}(j^{1}\avf)) & =\int_{\bdry\body}\soc(j^{1}\avf)-\int_{\body}\dee(p_{\st}(\diver\nhs)(\avf))+\int_{\body}\diver(\diver\nhs)(\avf),\\
 & =\int_{\bdry\body}\soc(j^{1}\avf)-\int_{\bdry\body}p_{\st}(\diver\nhs)(\avf)+\int_{\body}\diver(\diver\nhs)(\avf).
\end{split}
\end{equation}
We note that in the first integral $\soc$ is a section of $\L{J^{1}\avb,\ext^{n-1}T^{*}\spc}$,
and so it plays the role of a variational stress on the $(n-1)$-dimensional
manifold $\bdry\body$. One may therefore use the definition of the
generalized divergence to obtain
\begin{equation}
\soc(j^{1}\avf)=\dee(p_{\st}(\soc)(\avf))-\diver\soc(\avf)
\end{equation}
where $p_{\st}(\soc)$ is a section of $\L{\avb,\ext^{n-2}T^{*}\bdry\body}$
and $\diver\soc$ is a section of $\L{\avb,\ext^{n-1}T^{*}\bdry\body}$.
In other words, $p_{\st}(\soc)$ is a surface stress as one would
expect in second order continuum mechanics. We conclude that
\begin{equation}
\begin{split}\int_{\body}\nhs(j^{1}(j^{1}\avf)) & =\int_{\bdry\body}\dee(p_{\st}(\soc)(\avf))-\int_{\bdry\body}\diver\soc(\avf)\\
 & \qquad\qquad\qquad-\int_{\bdry\body}p_{\st}(\diver\nhs)(\avf)+\int_{\body}\diver(\diver\nhs)(\avf),\\
 & =\int_{\bdry(\bdry\body)}p_{\st}(\soc)(\avf)-\int_{\bdry\body}\diver\soc(\avf)\\
 & \qquad\qquad\qquad-\int_{\bdry\body}p_{\st}(\diver\nhs)(\avf)+\int_{\body}\diver(\diver\nhs)(\avf),
\end{split}
\end{equation}
As $\bdry(\bdry\body)=0$, it follows that
\begin{equation}
\int_{\body}\nhs(j^{1}(j^{1}\avf))=+\int_{\body}\diver(\diver\nhs)(\avf)-\int_{\bdry\body}(\diver p_{\st}(\nhs))(\avf)-\int_{\bdry\body}p_{\st}(\diver\nhs)(\avf).
\end{equation}

\newpage{}

\bigskip{}

\noindent \textbf{\textit{Acknowledgments.}} This work was partially
supported by the Pearlstone Center for Aeronautical Engineering Studies
at Ben-Gurion University.

\end{document}